\documentclass[prd,twocolumn,superscriptaddress,showpacs,nofootinbib,preprintnumbers]{revtex4}
\usepackage{amsmath}
\usepackage{amsfonts}
\usepackage{graphicx}
\usepackage{dcolumn}
\usepackage{hyperref}

\def\rd{{\rm d}}

\def\bea{\begin{eqnarray}}
\def\eea{\end{eqnarray}}
\def\beqa{\begin{equation}}
\def\eeqa{\end{equation}}

\def\be{\begin{equation}}
\def\ee{\end{equation}}

\def\5{\overline 5}
\def\vp{\varphi}

\newcommand{\tX}{\tilde{X}}
\begin{document}

\title{A unified approach to scaling solutions in a general \\
cosmological background}
\author{Shinji Tsujikawa}
\affiliation{Department of Physics, Gunma National College of
Technology, Gunma 371-8530, Japan}
\author{M.~Sami}
\affiliation{IUCAA, Post Bag 4, Ganeshkhind,
Pune 411 007, India}
\date{\today}
\vskip 1pc
\begin{abstract}

Our ignorance about the source of cosmic acceleration has stimulated study
of a wide range of models and modifications to gravity.
Cosmological scaling solutions in any of these theories are privileged
because they represent natural backgrounds relevant to dark energy.
We study scaling solutions in a generalized background
$H^2 \propto \rho_T^n$ in the presence of a scalar field
$\vp$ and a barotropic perfect fluid, where $H$ is a Hubble rate
and $\rho_T$ is a total energy density.
The condition for the existence of scaling solutions restricts
the form of Lagrangian to be $p=X^{1/n}g(Xe^{n\lambda \vp})$,
where $X=-g^{\mu\nu} \partial_\mu \vp
\partial_\nu \vp /2$ and $g$ is an arbitrary function.
This is very useful to find out scaling solutions
and corresponding scalar-field potentials in a broad class
of dark energy models including (coupled)-quintessence,
ghost-type scalar field, tachyon and k-essence.
We analytically derive the scalar-field equation of
state $w_\vp$ and the fractional density $\Omega_\vp$ and
apply it to a number of dark energy models.

\end{abstract}

\pacs{98.70.Vc, 98.80.Cq}

\maketitle 

\section{Introduction}

Accelerated expansion seems to have played an extremely 
important role in the dynamical history of our universe.
The inflationary paradigm at early epoch not only solves horizon and
flatness problems but predicts nearly scale-invariant
and adiabatic density perturbations consistent with temperature
anisotropies in Cosmic Microwave Background (CMB).
The late time acceleration is supported by observations of high
redshift type Ia supernovae and, more indirectly,
by observations of the CMB and galaxy clustering.

Within the framework of general relativity, cosmic acceleration can be
sourced by an energy-momentum tensor which has a large negative
pressure called dark energy (see Refs.~\cite{phiindustry} for review).
One of the well known candidates
of dark energy is provided by a cosmological constant.
Although the cosmological constant does not require an {\it adhoc}
assumption for its introduction, this suffers from
an extreme fine tuning problem due to its non-dynamical nature.
This problem can be alleviated in models of dynamically evolving dark 
energy
called quintessence \cite{Quin}.
In addition to quintessence a wide variety of scalar field dark energy
models have recently been proposed, including
k-essence \cite{Kesse,COY},
ghosts (phantoms) \cite{ghost} and Born-Infeld scalars (rolling
tachyon \cite{tachyon}, massive scalars \cite{massive},
phantom tachyons \cite{ptachyon}), with the
last one being originally motivated by string theory.

In order to obtain viable dark energy models, it is necessary that
the energy density of the scalar field remains
unimportant during most of the thermal history and
emerges only at late times to
account for the current acceleration of universe.
It is, therefore, important to investigate cosmological scenarios
in which the energy density of the scalar field mimics the background 
energy density. These solutions are called {\it scaling solutions} or
{\it trackers}. In this paper we use ``scaling solutions'' as a meaning 
that the energy density of the field  decreases proportionally to that of a
barotropic perfect fluid\footnote{We use ``trackers'' as a meaning that
the energy density of the field simply catches up that of the fluid.
}. Then the equation of state of the scalar field
equals to that of the fluid ($w_\vp=w_m$) for scaling solutions
in the absence of the coupling $Q$ between them.
In this case it is not possible to get an accelerated expansion at late
times
provided that the background fluid is dominated by a non-relativistic dark
matter
($w_m=0$). However the coupled quintessence scenario \cite{Amendola}
provides a possibility that scaling solutions give the acceleration of the
universe with a suitable fraction of dark energy ($\Omega_\vp \simeq 0.7$).

In General Relativity (GR) steep exponential potentials give rise to 
scaling solutions for a minimally coupled scalar field \cite{CLW} 
allowing the field energy density
to mimic the background being sub-dominant 
during radiation- and matter-dominant eras.
We can obtain the current accelerated expansion provided that 
the exponential potential becomes shallow to support the slow-roll 
at large values of the field \cite{Ed}.
Another interesting example is the model in which steep exponential 
potentials reduce to a particular power-law type at 
late times such that the universe exits from the scaling regime
(e.g., $V(\phi)=V_0\left[\cosh ({\alpha}
\phi/M_p)-1\right]^q,~q>0$ \cite{Sahni}). 
In coupled quintessence scenarios one can exploit scaling solutions
at late times as well when the coupling $Q$ grows during the transition 
to a scalar-field dominant era \cite{AT}.
Thus scaling solutions provide us very useful 
information for constructing dark energy models. 

The existence of scaling solutions has been extensively studied in a number
of cosmological scenarios, such as standard GR, braneworlds
[Randall-Sundrum (RS) and Gauss-Bonnet (GB)], and tachyon
\cite{LS}-\cite{SO}.
Nevertheless these works restrict the analysis to each different scenario.
In this paper we present a unified framework to investigate
scaling solutions in a general cosmological
background characterized by $H^2 \propto \rho_T^n$, where
$H$ is the Hubble rate and $\rho_T$ is the total energy density.
The GR, RS, GB cases correspond to $n=1$, $n=2$ and $n=2/3$,
respectively. Our formalism provides a generic method to study these
solutions for all the known scalar field systems like quintessence, tachyon, 
k-essence and ghost-type field.

We implement the coupling $Q$ between the field and the barotropic fluid
and obtain a general form of the Lagrangian for the existence of
scaling solutions, see Eq.~(\ref{scap}). Therefore
our analysis includes coupled quintessence scenario that
leads to an accelerated expansion even when the background fluid is
dominated by non-relativistic dark matter.
Our algorithm automatically generates scalar-field potentials which
give rise to scaling solutions in a general cosmological background. We
recover the already known solutions
in a generic way and also find new solutions in the presence of
the coupling $Q$. We also derive the equation of state $w_\vp$ and
the fractional density $\Omega_\vp$ for the field $\vp$.

\section{The lagrangian for scaling solutions}

Let us consider the following general 4-dimensional action
\begin{eqnarray}
\label{action}
{\cal S}=
\int d^4 x \sqrt{-g} \left[\frac{M_P^2}{2}\, R
+ p(X, \varphi)\right]+{\cal S}_m [\varphi, \Psi_i, g_{\mu \nu}],
\end{eqnarray}
where $R$ is a scalar curvature, $\vp$ is a scalar field
with $X$ defined as
$X \equiv -g^{\mu\nu} \partial_\mu \varphi
\partial_\nu \varphi /2$, and
$p(X, \vp)$ is a scalar-field Lagrangian that is a function in
terms of $X$ and $\vp$.
${\cal S}_m$ is an action for matter fields $\Psi_i$,
which is generally dependent on $\vp$ as well.
Hereafter we set the Planck mass $M_p$ to be unity.

We shall study cosmological scaling solutions
in a spatially flat Friedmann-Robertson-Walker (FRW)
background spacetime:
\begin{equation}
\label{friedman}
ds^2=-dt^2+a^2(t)d{\bf x}^2\,,
\end{equation}
where $a(t)$ is a scale factor.
We consider an effective Friedmann equation which is given by
\be 
\label{Hubble}
H^2=\beta_n^2 \rho_T^n\,,
\ee
where $\beta_n$ and $n$ are constants, and
$\rho_T$ is a total energy density of the universe.
For the background (\ref{friedman}) the equation
for $\varphi$ is \cite{PT}
\begin{equation}
\label{geneeq}
{\ddot \vp}\left(p_X + \dot \vp ^2 p_{XX} \right)
+ 3 H p_X \, \dot \vp
+ 2Xp_{X \vp} 
- p_\vp = -\sigma\,,
\end{equation}
where a suffix $X$ or $\vp$ denotes a partial derivative with
respect to $X$ or $\vp$, respectively.
Here the scalar charge $\sigma$ corresponds to the coupling
between a matter and the field $\vp$, which
is defined by the relation
$\delta {\cal S}_m/\delta \vp = - \sqrt{-g}\, \sigma$.
We consider a cosmological scenario in which the universe
is filled by the scalar field $\varphi$ and by one type of
barotropic perfect fluid with an equation
of state $w_m=p_m/\rho_m$.
Rewriting Eq.~(\ref{geneeq}) in terms of the energy density,
$\rho=2Xp_X-p$, of the scalar field, we get
\begin{eqnarray}
\label{geneeq1}
\frac{\rd \rho}{\rd N}+3(1+w_\vp)\rho
=-Q\rho_m \frac{\rd \vp}{\rd N}\,,
\end{eqnarray}
where 
\begin{equation}
\label{def}
N \equiv {\rm ln}\,a\,,~~~
w_{\vp} \equiv p/\rho\,,~~~
Q(\vp) \equiv \sigma /\rho_m\,.
\end{equation}
The energy density $\rho_m$ of the fluid satisfies
\begin{eqnarray}
\label{geneeq2}
 \frac{\rd \rho_m}{\rd N} + 3(1+w_m) \rho_m =  Q \rho_m
 \frac{\rd \vp}{\rd N}\,.
\end{eqnarray}
We define the fractional densities of $\rho$ and $\rho_m$ as
\begin{eqnarray}
\label{Omedef}
\Omega_\varphi \equiv \frac{\rho}{(H/\beta_n)^{2/n}}\,,~~~~
\Omega_m \equiv \frac{\rho_m}{(H/\beta_n)^{2/n}}\,,
\end{eqnarray}
which satisfy $\Omega_\vp+ \Omega_m =1$ by Eq.~(\ref{Hubble}).

We are interested in asymptotic scaling solutions where both
the fractional density  $\Omega_\vp$ and the equation
of state parameter $w_\vp$
are constant, which gives $\rho/\rho_m={\rm const}$.
This translates into the condition
${\rm d} \log \rho/{\rm d}N = {\rm d}\log \rho_m/{\rm d}N $.
Assuming that $Q$ is not a time-varying function
in the scaling regime we get the following relation
from Eqs.~(\ref{geneeq1}) and (\ref{geneeq2}):
\begin{equation} 
\label{dphi}
\frac{\rd \vp}{\rd N} = \frac{3\Omega_\vp}{Q}
(w_m - w_\vp) = {\rm const.}
\end{equation}
Then we find the scaling
behavior of $\rho$ and $\rho_m$:
\begin{equation} 
\label{sca}
\frac{\rd {\rm log} \rho}{\rd N}=
\frac{\rd {\rm log} \rho_m}{\rd N}=
-3(1+w_s)\,,
\end{equation}
where the effective equation of state is
\begin{equation} 
\label{ws}
w_s \equiv w_m+\Omega_\vp (w_\vp-w_m)\,.
\end{equation}
Therefore $\rho$ and $\rho_m$
do not scale according to $w_\vp$ and $w_m$
in the presence of the coupling $Q$.

{}From the definition of $X$ one finds (in the FRW background)
\begin{equation} 
\label{Xdef}
2 X = H^2 \left(\frac{\rd \vp}{\rd N}\right)^2 \propto \, H^2
\propto \rho_T^n\,,
\end{equation}
which shows that the scaling property of $X$ is the same as $\rho^n$
and $\rho_m^n$. Then we obtain
\begin{equation} 
\label{X2eq}
\frac{\rd {\rm log} X}{\rd N}=-3n(1+w_s)  \,.
\end{equation}
Since $p=w_\vp \rho$ scales in the same way as $\rho$,
one has $\rd {\rm log}\,p/\rd N=-3(1+w_s)$.
Here the pressure density $p$ corresponds to the Lagrangian of
the scalar field and is a function of $X$ and $\vp$.
Therefore we obtain the following relation
by using Eqs.~(\ref{dphi}) and (\ref{X2eq}):
\begin{equation}
\label{pform}
n\frac{\partial \log p}{\partial \log X} -
\frac{1}{\lambda} \frac{\partial \log p}{\partial \vp} = 1\,,
\end{equation}
where 
\begin{equation}
\label{lam}
\lambda\,  \equiv\,
Q \frac{1+w_m - \Omega_\vp (w_m - w_\vp)}
{\Omega_\vp (w_m-w_\vp)}\,.
\end{equation}
This equation gives a constraint on the functional
form of $p(X, \vp)$ for the existence of scaling solutions:
\begin{equation} 
\label{scap}
p(X, \vp) = X^{1/n}\,
g\left(X e^{n\lambda \vp}\right)\,,
\end{equation}
where $g$ is any function in terms of $ Y \equiv 
X e^{n\lambda \vp}$. This coincides with what was obtained 
in Ref.~\cite{PT} in the GR case ($n=1$). 
One can easily show that $Y$ is constant
along the scaling solution, i.e.,
\begin{equation} 
X e^{n\lambda \vp} = Y_0
={\rm const}\,.
\end{equation}
This property tells us that $p$ is proportional to $X^{1/n}$
by Eq.~(\ref{scap}). This could be a defining property of scaling solutions
which means that the Lagrangian or the pressure density 
depends upon the kinetic energy alone in the scaling regime. 
For an ordinary scalar field it leads to constancy of the ratio of 
kinetic to potential energy which is 
often taken to be a definition of scaling solutions.

{}From the pressure density (\ref{scap}) we obtain
the energy density $\rho$ as $\rho=X^{1/n}(2/n-1+2Yg'/g)g$,
where a prime denotes a derivative
in terms of $Y$. Then the equation of state $w_\vp=p/\rho$ reads
\begin{equation} 
\label{wp}
w_\vp=\left(\frac{2}{n}-1+2\alpha\right)^{-1}\,,
\end{equation}
where 
\begin{equation} 
\label{alp}
\alpha \equiv \, \left. \frac{{\rm d} \log g(Y)}
{{\rm d} \log Y} 
\right|_{Y = Y_0}\,.
\end{equation} 
Making use of Eqs.~(\ref{dphi}), (\ref{Xdef}) and
(\ref{lam}), we get
\begin{equation}
3H^2=\frac{2(Q+\lambda)^2}{3(1+w_m)^2}X\,.
\end{equation} 
Then the fractional density (\ref{Omedef}) of
the field $\vp$ yields
\begin{equation} 
\label{Omevp}
\Omega_\vp=\left(\frac{9\beta_n^2(1+w_m)^2}
{2(Q+\lambda)^2}\right)^{1/n}\frac{g(Y_0)}{w_\vp}\,.
\end{equation} 

By combining  Eq.~(\ref{dphi}) with Eq.~(\ref{Omedef})
together with the relation $w_\vp=p/\rho$,
we find that $g$ in
Eq.~(\ref{scap}) can be written as
\begin{equation}
\label{gY0}
g(Y_0)=-Q\left(\frac{2}{9\beta_n^2}\right)^{1/n}
\frac{w_\vp}{w_\vp- w_m}\left(\frac{1+w_m}
{Q+\lambda}\right)^{(n-2)/n}\,.
\end{equation} 
Then Eq.~(\ref{Omevp}) yields
\begin{equation}
\label{Omephi}
\Omega_\vp=\frac{Q}{Q+\lambda}
\frac{1+w_m}{w_m-w_\vp}\,.
\end{equation} 
Once the functional form of $g(Y)$ is known, the equation of
state $w_\vp$ is determined by Eq.~(\ref{wp})
with Eq.~(\ref{alp}).
Then we get the fractional density $\Omega_\vp$
from Eq.~(\ref{Omephi}).

For scaling solutions an acceleration parameter yields
\begin{equation}
\label{q}
-q \equiv \frac{\ddot{a}a}{\dot{a}^2}=
1-\frac{3n(1+w_m)\lambda}{2(\lambda+Q)}\,.
\end{equation} 
When $Q=0$ the condition $-q>0$ gives $w_m<2/(3n)-1$.
For example $w_m<-1/3$ for $n=1$.
For non-relativistic dark matter ($w_m=0$), an accelerated expansion
occurs only for $n<2/3$.
If we account for the coupling $Q$, it is possible to
get an acceleration even for $n \ge 2/3$.
The condition for acceleration corresponds to
\begin{equation}
\label{acc}
\frac{Q}{\lambda}>\frac{3n(1+w_m)-2}{2}\,,
\end{equation} 
which is useful for the construction of realistic dark
energy scenarios.

\section{Application to dark energy models}

\subsection{Ordinary scalar fields}

Let us first obtain the form of lagrangian when $p$
is written in the form
\begin{equation} 
\label{sum}
p(X, \vp) = f(X)-V(\vp)\,.
\end{equation}
Then by Eq.~(\ref{pform}) one gets
\begin{equation} 
\label{sumre}
nX \frac{\rd f}{\rd X}-f(X)=
-\frac{1}{\lambda}\frac{\rd V}{\rd \vp}-V
\equiv C \,,
\end{equation}
where $C$ is a constant.
Integrating this relation gives $f=c_1X^{1/n}-C$
and $V=c_2e^{-\lambda \vp}-C$, which restricts
the form of the Lagrangian to be
\begin{equation} 
p=c_1X^{1/n}-c_2e^{-\lambda \vp}\,.
\end{equation}
In the case of GR ($n=1$), this corresponds to a standard
canonical scalar field with an exponential
potential \cite{CLW}.
When $n \ne 1$ the Lagrangian does not take a canonical
form, so the exponential potential does not correspond
to scaling solutions.

One may look for scaling solutions that give a standard kinematic
term by an appropriate transformation to a new variable in
Eq.~(\ref{scap}).
Introducing a new variable $\phi \equiv e^{\beta \lambda \vp}$,
we obtain $Y_0=\tilde{X}\phi^{(n-2\beta)/\beta}/
\beta^2\lambda^2={\rm const}$, where $\tilde{X} \equiv
-g^{\mu\nu} \partial_\mu \phi
\partial_\nu \phi /2$.
Then the Lagrangian (\ref{scap}) can be written as
\begin{equation} 
\label{lagor}
p=\frac{Y_0^{1/n}}{\phi^{1/\beta}}g(Y_0)
=Y_0^{1/n} \left(\frac{\tX}{\beta^2 \lambda^2
Y_0}\right)^{1/(n-2\beta)}g(Y_0)\,.
\end{equation}
Since $p$ is proportional to $\tX^{1/(n-2\beta)}$, the transformation that
gives $p \propto \tX$ corresponds to $\beta=(n-1)/2$, i.e.,
$\phi=e^{(n-1)\lambda \vp/2}$.
In this case we have $p \propto \phi^{-2/(n-1)}$ by Eq.\,(\ref{lagor}),
which means that the potential of the field $\phi$ corresponding to
scaling solutions is
\begin{equation} 
\label{poten}
V(\phi)=V_0\phi^{-2/(n-1)}\,,
\end{equation}
where $V_0$ is a constant.
For example one has an inverse square potential
$V(\phi)=V_0\phi^{-2}$ for $n=2$, which agrees with
what was obtained in Ref.~\cite{MLC}.
The Gauss-Bonnet braneworld ($n=2/3$) gives the potential
$V(\phi)=V_0\phi^6$, as shown in Ref.~\cite{SST}.

If we choose the function $g(Y)$ as
\begin{equation}
\label{gY1} 
g(Y)=c_1Y^{1-1/n}-c_2Y^{-1/n}\,,
\end{equation}
then we get the Lagrangian
\begin{eqnarray}
\label{p}
p&=& c_1X e^{(n-1)\lambda \vp}-c_2e^{-\lambda \vp} \\
  &=& \frac{4c_1}{(n-1)^2\lambda^2}\tilde{X}
  -c_2\phi^{-2/(n-1)}\,,
\label{p2}
\end{eqnarray}
where the last equality is valid for $n \ne 1$.
Therefore the choice (\ref{gY1}) gives the canonical
Lagrangian with potential given by (\ref{poten}) for
$n \ne 1$. When $n=1$ the lagrangian (\ref{p}) itself is
canonical. 
We can obtain $w_\vp$ and $\Omega_\vp$ in these
cases by using the function (\ref{gY1}).
Note that a normal scalar field corresponds to
$\epsilon \equiv 4c_1/((n-1)^2\lambda^2)>0$, whereas
a ghost field to $\epsilon<0$.

\subsubsection{The GR case}

Let us consider the GR case ($n=1$).
Since $c_2/Y_0=c_1-g(Y_0)$ by Eq.~(\ref{gY1}),
one has $\alpha=c_1/g(Y_0)-1$ by using Eq.~(\ref{alp}).
{}From Eq.~(\ref{gY0}) we find that $g(Y_0)$ can be written as
\begin{equation}
\label{gYGR} 
g(Y_0)=\frac{1}{1+w_m}
\left[2c_1w_m-\frac{2Q(Q+\lambda)}
{3(1+w_m)} \right]\,.
\end{equation}
Then we obtain the equation of state
\begin{equation}
\label{OmeGR}
w_\vp=\frac{3c_1w_m(1+w_m)-Q(Q+\lambda)}
{3c_1(1+w_m)+Q(Q+\lambda)}\,.
\end{equation}
Inserting this into Eq.~(\ref{Omephi}) gives
\begin{equation}
\label{omeGR}
\Omega_\vp=\frac{3c_1(1+w_m)+Q(Q+\lambda)}
{(Q+\lambda)^2}\,.
\end{equation}

When $Q=0$ the above results reduce to $w_\vp=w_m$
and $\Omega_\vp=3c_1(1+w_m)/\lambda^2$.
This coincides with the scaling solution for an exponential potential
obtained in Ref.~\cite{CLW}.
In this case although the energy density of the field $\vp$
contributes to some portion of the total energy density,
we can not obtain an acceleration of the universe for  a
normal fluid satisfying $w_\vp > -1/3$.
However the presence of the coupling $Q$ opens up a
possibility of an accelerated expansion.
In the case of non-relativistic dark matter
($w_m=0$), we obtain $w_\vp=-Q(Q+\lambda)
/(3c_1+Q(Q+\lambda))$ and
$\Omega_\vp=(3c_1+Q(Q+\lambda))/(Q+\lambda)^2$,
which agrees with the coupled quintessence scenario
in Ref.~\cite{Amendola}.

\subsubsection{The RS case}

In the RS case ($n=2$) we have $c_2/\sqrt{Y_0}=
c_1\sqrt{Y_0}-g(Y_0)$, $\alpha=-1/2+c_1\sqrt{Y_0}/g(Y_0)$
and $w_\vp=g(Y_0)/(2c_1\sqrt{Y_0})$.
Hereafter we shall use the parameter $\epsilon=4c_1/\lambda^2$
that is positive for an ordinary scalar field.
Then by Eq.~(\ref{gY0}) one gets
\begin{equation} 
g(Y_0)=\frac{\epsilon \lambda^2}{2}
\sqrt{Y_0}w_m-\frac{\sqrt{2}Q}
{3\beta_2}\,,
\end{equation}
where
\begin{equation}
\frac{\epsilon \lambda^2}{2} \sqrt{Y_0} = 
\frac{-\sqrt{2}Q/(3\beta_2)
+ \sqrt{2Q^2/(9\beta_2^2)+\epsilon
\lambda^2c_2(1-2w_m)}}{1-2w_m}\,.
\end{equation}
Here the conditions $\epsilon>0$ and $c_2>0$
are assumed.

The equation of state for the field $\vp$ is
\begin{equation} 
w_\vp=w_m-\frac{(1-2w_m)Q}
{-Q + \sqrt{Q^2+(9/2)\beta_2^2 \epsilon \lambda^2
c_2(1-2w_m)}}\,.
\end{equation}
By Eq.~(\ref{Omevp}) we find
\begin{equation}
\Omega_\vp=\frac{1+w_m}{1-2w_m}
\frac{-Q + \sqrt{Q^2+(9/2)\beta_2^2 \epsilon \lambda^2
c_2(1-2w_m)}}{|Q+\lambda|} \,.
\end{equation}
When $Q=0$ one has $w_\vp=w_m$ with a nonzero
value of $\Omega_\vp$, as is similar to the case of
$n=1$. If we include the coupling $Q$, we have
$w_\vp=-Q/(\sqrt{Q^2+(9/2)\beta_2^2\epsilon
\lambda^2 c_2}-Q)$ and $\Omega_\vp=
(\sqrt{Q^2+(9/2)\beta_2^2\epsilon \lambda^2 c_2}-Q)/
|Q+\lambda|$ for $w_m=0$.
Therefore it is possible to have an accelerated expansion
in the presence of the coupling $Q$.

\subsection{Phantoms and ghost condensates}

A ghost (phantom) scalar field corresponds to a negative
sign of $c_1$ in Eq.~(\ref{gY1}).
In the GR case with $Q=0$ one has
$\Omega_\vp=3c_1(1+w_m)/\lambda^2<0$
for $w_m>-1$, which means the absence of
viable scaling solutions. In the presence of the coupling $Q$
there exist scaling solutions that satisfy
the condition for an accelerated expansion,
see Eqs.~(\ref{OmeGR}) and (\ref{omeGR}).
In the RS case since $g(Y_0)/w_\vp=2c_1\sqrt{Y_0}$
is negative in Eq.~(\ref{Omevp}) for $c_1<0$, one can not
obtain viable scaling solutions.

We need to keep in mind that phantoms are generally plagued
by severe ultraviolet quantum instabilities \cite{Cline}.
However it was shown in Ref.~\cite{Arkani} that
a scalar field with a negative sign kinematic term does not
necessarily lead to inconsistencies, provided that a suitable
structure of higher-order kinematics terms are present in the
effective theory. Let us consider the Lagrangian of the form
\begin{equation} 
\label{ghost}
p=\epsilon X+ce^{\lambda \vp}X^2\,,
\end{equation}
where negative $\epsilon$ corresponds to the phantom.
This is motivated by dilatonic higher-order corrections
in low energy effective string theory \cite{PT}.
Since the function $g(Y)$ is $g(Y)=\epsilon+cY$ in the GR
case ($n=1$), we obtain $\alpha=cY_0/(\epsilon+cY_0)$,
$w_\vp=(\epsilon+cY_0)/(\epsilon+3cY_0)$ and
\begin{equation} 
cY_0=-\frac{2Q(Q+\lambda)+3\epsilon(1-w_m^2)}
{3(1+w_m)(1-3w_m)}\,.
\end{equation}
Then we get
\begin{eqnarray} 
\label{ghostw1}
w_\vp &=& \frac{3\epsilon(1+w_m)w_m+Q(Q+\lambda)}
{3\epsilon(1+w_m)+3Q(Q+\lambda)}
\,, \\
\Omega_\vp &=& \frac{3(1+w_m)\left[-\epsilon (1+w_m)-Q(Q+\lambda)
\right]}{(Q+\lambda)^2(1-3w_m)}\,.
\label{ghostw2}
\end{eqnarray}
This agrees with those obtained in Ref.~\cite{PT}
for $w_m=0$ and $\epsilon=-1$.
The stability of quantum fluctuations is ensured for
$p_X+2Xp_{XX} \ge 0$ and $p_X \ge 0$, which
corresponds to the condition $cY_0 \ge 1/2$ in our case.
This translates into the relation $Q(Q+\lambda) \le 3(1+w_m)^2/4$.
On the other hand the condition for acceleration requires
$Q/\lambda>(1+3w_m)/2$. The values of the coupling $Q$
satisfying both of these conditions
provide viable scaling solutions.

\subsection{Tachyon fields}

The Lagrangian for a tachyon field is
given by \cite{tachyon}
\begin{equation} 
\label{tachlag}
p=-V(\phi) \sqrt{1-\dot{\phi}^2}\,.
\end{equation}
Apparently the general form (\ref{scap}) does not seem
to include this case, but one can rewrite the Lagrangian
(\ref{scap}) by introducing a new field
$\phi=e^{\beta \lambda \vp}/(\beta \lambda)$.
Since the quantity $Y$ is written as
$Y=\tX (\beta \lambda \phi)^{n/\beta-2}$ with
$\tilde{X} \equiv -g^{\mu\nu} \partial_\mu \phi
\partial_\nu \phi /2$ under this
transformation, we obtain $Y=\tX$ for $\beta=n/2$.
Then the Lagrangian (\ref{scap}) yields
\begin{equation} 
\label{tachlag2}
p=\left(\frac{n \lambda \phi}{2}\right)^{-2/n}
\tX^{1/n}g(\tX)\,.
\end{equation}
This is a system $p(\tX, \phi)=V(\phi)f(\tX)$
with potential
\begin{equation}
\label{potacy} 
V(\phi)=V_0\phi^{-2/n}\,,
\end{equation}
and $f(\tX)=\tX^{1/n}g(\tX)$.

We get the tachyon system (\ref{tachlag}) by the choice
\begin{equation} 
\label{gYtach}
g(Y)=-cY^{-1/n} \sqrt{1-2\epsilon Y},~~
{\rm with}~~\epsilon=1\,,
\end{equation}
where $c$ is positive.
We introduce a parameter $\epsilon$ so that
the system includes a phantom tachyon field with
$\epsilon=-1$ \cite{ptachyon}.
By Eq.~(\ref{potacy}) one has an inverse square potential
$V(\phi)=V_0\phi^{-2}$
for $n=1$ (GR) for the existence of scaling solutions \cite{AL}.
We have $V(\phi)=V_0\phi^{-1}$ for $n=2$ (RS) and
$V(\phi)=V_0\phi^{-3}$ for $n=2/3$ (GB).
Hereafter we shall derive $w_\vp$ and $\Omega_\vp$
for the function (\ref{gYtach}) in GR and RS cases.

\subsubsection{The GR case}

When $n=1$ we obtain $g(Y_0)=-c\sqrt{1-2\epsilon Y_0}/Y_0$,
$\alpha=(\epsilon Y_0-1)/(1-2\epsilon Y_0)$ and
$w_\vp=2\epsilon Y_0-1$.
Since $1-2\epsilon Y_0 \ge 0$ the equation of state ranges
in $w_\vp \le 0$. 
Making use of Eq.~(\ref{gY0})
and $g(Y_0)=-2c\sqrt{-w_\vp}/(w_\vp+1)$, one gets the
following equation 
\begin{equation} 
\label{relaGR}
\frac{x^2+w_m}{x(1-x^2)}=
\frac{Q(Q+\lambda)}{3c\epsilon(1+w_m)}\,,
\end{equation}
where $x \equiv \sqrt{-w_\vp}$.
The solution in the limit $Q \to 0$ corresponds to
$x=\sqrt{-w_m}$, i.e., $w_\vp=w_m$.
Note that the existence of scaling solutions requires
the condition $w_m<0$, as was pointed out in
Ref.~\cite{AL}.

One can approximately obtain the solution for Eq.~(\ref{relaGR})
when the coupling $Q$ is small.
Let us write the solution as $x=\sqrt{-w_m}+\delta$, where $\delta$ is
small relative to $\sqrt{-w_m}$.
Substituting this for Eq.~(\ref{relaGR}),
we find that $\delta$ is given by $\delta=\lambda Q/(6c\epsilon)$.
Then we get
\begin{eqnarray} 
w_\vp &=& w_m-\frac{\lambda \sqrt{-w_m}}
{3c\epsilon}Q\,, \\
\Omega_\vp &=& \frac{3c\epsilon}{(Q+\lambda)\lambda}
\frac{1+w_m}{\sqrt{-w_m}}\,,
\end{eqnarray}
which are valid when the coupling $Q$ is small.
In the limit $Q \to 0$ one has
$\Omega_\vp=3c\epsilon(1+w_m)/(\lambda^2\sqrt{-w_m})$,
which agrees with the result in Ref.~\cite{AL}
for $\epsilon=1$.
The phantom tachyon ($\epsilon=-1$) corresponds to $\Omega_\vp<0$
for $-1<w_m<0$ under the acceleration condition (\ref{acc}),
which means that viable scaling solutions do not exist.

Even if scaling solutions do not exist for a fluid satisfying $w_m \ge 0$,
this does not mean that we do not have a plausible dark energy scenario
in the tachyon system.
In fact there is a stable critical point that approaches
$\Omega_\vp=1$ and $\Omega_m=0$ with an accelerated expansion
for the potential (\ref{potacy}) \cite{AL}.
One can construct a viable dark energy model that evolves toward this
critical point in the future with the present value $\Omega_\vp \sim 0.7$
provided that initial conditions of the field
are appropriately chosen \cite{BJP}.
In addition an inverse power-law potential $V(\phi) \propto \phi^{-q}$
with $q<2$ leads to an acceleration of the universe at late times, while it
does not for $q>2$ \cite{AF}.
Thus the potential corresponding to scaling solutions
marks the border between accelerated and decelerated expansions, which
provides a useful information for the construction of dark energy models.

\subsubsection{The RS case}

When $n=2$ we have $g(Y_0)=-c\sqrt{1/Y_0-2\epsilon}$,
$\alpha=-1/(2(1-2\epsilon Y_0))$ and $w_\vp=1-1/(2\epsilon Y_0)$,
which means that $g(Y_0)=-c\sqrt{-2\epsilon w_\vp}$.
Then $w_\vp<0$ for $\epsilon>0$ and $w_\vp>0$ for $\epsilon<0$.
By using these relations together with Eq.~(\ref{gY0}),
we find that 
\begin{eqnarray} 
\frac{x^2+\epsilon w_m}{x}=\frac{Q}{3c\beta_2}\,,
\end{eqnarray}
where $x \equiv \sqrt{-\epsilon w_\vp}$.
In this case one can obtain $w_\vp$ and $\Omega_\vp$
for any values of $Q$, i.e.,
\begin{equation} 
\label{tacRS1}
w_\vp = -\frac{1}{4\epsilon} \left[\frac{Q}{3c\beta_2}
+\sqrt{\frac{Q^2}{9c^2\beta_2^2}-4\epsilon w_m}
\right]^2\,,
\end{equation}
\begin{equation} 
\Omega_\vp = 6\epsilon c\beta_2 \left| \frac{1+w_m}
{Q+\lambda}\right| 
\left[\frac{Q}{3c\beta_2}
+\sqrt{\frac{Q^2}{9c^2\beta_2^2}-4\epsilon w_m}
\right]^{-1}. 
\label{tacRS2}
\end{equation}
When $\epsilon=1$ and $Q=0$ scaling solutions
exist only for $w_m<0$, but the presence of the coupling
$Q$ allows a possibility of their existence even for $w_m>0$.
When $\epsilon=-1$ one has $w_\vp>0$
and $\Omega_\vp<0$ by Eqs.~(\ref{tacRS1}) and (\ref{tacRS2}),
which means the absence of ideal scaling solutions.

\subsection{K-essence}

K-essence scenario is characterized by the
pressure density \cite{Kesse}
\begin{eqnarray}
\label{Kesse} 
p=V(\phi)f(\tilde{X})\,,
\end{eqnarray}
where $\tilde{X}=-g^{\mu\nu} \partial_\mu \phi
\partial_\nu \phi /2$.
Since the Lagrangian (\ref{scap}) can be written as a decoupled form
(\ref{tachlag}) under a transformation $\phi=2e^{n\lambda \vp/2}/(n
\lambda)$, the k-essence Lagrangian (\ref{Kesse}) has
a scaling solution for
\begin{eqnarray}
\label{Kessecon} 
V(\phi) \propto \phi^{-2/n}\,,~~{\rm and}~~
\tilde{X}={\rm const}\,.
\end{eqnarray}
Note that the function $f(\tilde{X})$ can be chosen arbitrary provided that
the conditions (\ref{Kessecon}) are satisfied.
These conditions mean that the field $\phi$ evolves with a constant 
velocity
along the potential $V(\phi)=V_0\phi^{-2/n}$.
This is a general property of scaling solutions in k-essence scenario.

The pressure density of the form,
\begin{eqnarray}
p(X, \vp)=K(\vp)X+L(\vp)X^2\,,
\end{eqnarray}
is transformed to the Lagrangian (\ref{Kesse})
with $V(\phi)=K^2/L$ and $f(\tX)=-\tX+\tX^2$
by the field redefinitions \cite{COY}:
\begin{eqnarray}
\phi=\int^{\vp} {\rm d}\vp \sqrt{\frac{L}{|K|}}\,,~~~
\tX=\frac{L}{|K|}X\,.
\end{eqnarray}
Therefore the dilatonic ghost condensate considered in
Sec.\,III\,B belongs to a class of k-essence.
In fact the Lagrangian (\ref{ghost}) corresponds to
$K=\epsilon$ and $L=ce^{\lambda \vp}$, which gives
$\phi \propto e^{\lambda \vp/2}$ and
$V \propto e^{-\lambda \vp} \propto \phi^{-2}$.
Therefore scaling solutions (\ref{ghostw1}) and (\ref{ghostw2})
can be viewed as the system (\ref{Kessecon}) with $n=1$.

\vskip 1pc 
In this section we dealt with a variety of dynamical systems in
GR and RS backgrounds; a comment on the GB dynamics
is in order. We find
that the fundamental features of scaling solutions which appear in GR and
RS cases also persist in the GB background.
Since the algebra gets merely cumbersome, we have not shown
the results in the GB case.

\section{Conclusions}

In this paper we discussed cosmological scaling solutions
in a general cosmological background $H^2 \propto \rho_T^n$
including General Relativity,
Randall-Sundrum braneworld and Gauss-Bonnet braneworld.
The condition for the existence of scaling solutions restricts the
form of the Lagrangian to be Eq.~(\ref{scap}).
Since the starting action (\ref{action}) is very general, the formula
(\ref{scap}) is applicable for a wide variety of dark energy models
such as (coupled)-quintessence,
ghost-type scalar field, tachyon and k-essence.
This is a powerful tool to find out scaling solutions and corresponding
effective potentials in {\it any} scalar-field system.

We analytically derived the scalar-field equation of state $w_\vp$ and
the fractional density $\Omega_\vp$
in general, see Eqs.\,(\ref{wp}) and (\ref{Omephi})
with (\ref{alp}). We applied these formula to a number of dark energy
models and discussed the existence of viable scaling solutions.
In the absence of the coupling $Q$ between a scalar field and a
perfect barotropic fluid, it is not possible to get an acceleration
of the universe since the energy density of the field $\vp$
decreases in proportional to that of the background fluid for
scaling solutions. However the presence of the coupling
$Q$ allows to have an accelerated expansion
with an effective equation of state given by (\ref{ws}).

We have reproduced previous results about the form of the scalar field
potential in the GR case ($n=1$)
in the context of quintessence and tachyon.
We derived the potentials (\ref{poten}) and (\ref{potacy})
for scaling solutions from the Lagrangian (\ref{scap})
by redefining a new field $\phi$.
These may be obtained by considering the condition for power-law inflation
together with slow-roll parameters \cite{SST}, but out treatment
is more general since we did not use any approximations.

We also applied our formula to a ghost-type scalar field and
k-essence. We accounted for dilatonic higher-order terms
$e^{\lambda \vp}X^2$
in order to avoid severe ultraviolet instabilities present for a phantom
field
with a negative sign of $X$.
This scenario is closely linked with k-essence,
since the Lagrangian (\ref{ghost}) is transformed to (\ref{Kesse}) by a
field redefinition. By using Eq.~(\ref{scap}), we showed that
the Lagrangian (\ref{Kesse}) has a scaling solution when
$V(\phi) \propto \phi^{-2/n}$ and $\tilde{X}={\rm const}$.
Provided that these conditions are satisfied, scaling solutions exist
for any arbitrary function $f(\tilde{X})$ in Eq.~(\ref{Kesse})
including the tachyon case.

If the scalar-field potential is not steeper than
the one for scaling solutions, this leads to an accelerated expansion
since scaling solutions give the border of acceleration and deceleration.
Therefore our formalism is useful to construct realistic dark energy 
models.
It is also of interest to place constraints on scalar-field
potentials using supernova and CMB data and discriminate between a host of
dark energy models from future high-precision observations.

\section*{ACKNOWLEDGEMENTS}
We thank Edmund Copeland, Shuntaro Mizuno
and Federico Piazza for fruitful discussions and
Bruce A. Bassett  and Varun Sahni for useful comments and 
conversations.




\end{document}